\documentclass[useAMS, usenatbib]{mn2e}

\usepackage{psfig, epsf, epsfig, graphicx, epstopdf, amssymb, tabularx}
\usepackage{natbib}

\title[Formation of ultra-compact dwarfs]{Formation of 
ultra-compact dwarf galaxies from
supergiant molecular clouds}
\author[Morgan  Goodman and Kenji Bekki]
{Morgan Goodman${}^1$\thanks{E-mail:
morgan.goodman@uwa.edu.au}
and Kenji Bekki  \\ 
${}^1$ICRAR M468
The University of Western Australia
35 Stirling Hwy, Crawley
Western Australia, 6009 \\ }

\begin{document}

\date{Accepted, Received 2005 February 20; in original form }

\pagerange{\pageref{firstpage}--\pageref{lastpage}} \pubyear{2005}

\maketitle

\raggedbottom

\label{firstpage}

\begin{abstract}
The origin of ultra-compact dwarf galaxies (UCDs) is not yet clear. One possible formation path of UCDs is the threshing of a nucleated elliptical dwarf galaxy (dE, N), however, it remains unclear how such massive nuclear stellar systems were formed in dwarf galaxies. To better establish the early history of UCDs, we investigate the formation of UCD progenitor clusters from super giant molecular clouds (SGMCs), using hydrodynamical simulations. In this study we focus on SGMCs with masses $10^{7} - 10^{8} \rm M_{\odot}$ that can form massive star clusters that display physical properties similar to UCDs. We find that the clusters have extended star formation histories with two phases, producing multiple distinct stellar populations, and that the star formation rate is dependent on the feedback effects of SNe and AGB stars. The later generations of stars formed in these clusters are more compact, leading to a clearly nested structure, and these stars will be more He-rich than those of the first generation, leading to a slight colour gradient. The simulated clusters demonstrate scaling relations between $R_{\rm eff}$ and $M$ and $\sigma_{v}$ and $M$ consistent with those observed in UCDs and strongly consistent with those of the original SGMC. We discuss whether SGMCs such as these can be formed through merging of self-gravitating molecular clouds in galaxies at high-\it z.
\end{abstract}


\section{Introduction}

Ultra-compact dwarf galaxies (UCDs) are stellar systems that bridge the divide between definitions of globular clusters and dwarf galaxies. They were first discovered in spectroscopic surveys of the Fornax galaxy cluster by \cite{drinkwater2000compact} and \cite{hilker1999central}. Since then, dedicated surveys have discovered hundreds of UCDs within the Fornax cluster \citep{mieske2004fornax, firth2007compact, firth2008compact} in other galaxy clusters, including Virgo \citep{hacsegan2005acs, jones2006discovery}, Coma \citep{price2009hst} and Abell 1689 \citep{mieske2004ultracompact}, and around isolated galaxies such as NGC4546 \citep{norris2011ubiquity}. UCDs are stellar systems with luminosities between -16 $\gtrsim$ $M_{V}$ $\gtrsim$ -10 $M_{V}$, surface brightnesses similar to those of globular clusters (19-23 mag arcsec$^{-2}$) and effective radii 10 $<$ R$_{\rm eff}$ $<$ 100 pc. They exhibit high dynamical masses for objects of this size, between 10$^{6}$ $M_{\odot}$ $\lesssim M_{\rm dyn} \lesssim 10^{8} M_{\odot}$ \citep{brodie2011relationships, mieske2008nature}, and may be some of the densest stellar systems known to exist. The dynamical masses of UCDs are in excess of those predicted by stellar population models, suggesting that some UCDs may be host to a central supermassive black hole \citep{mieske2013central, seth2014supermassive}. 

Although there are many observations of UCDs, in many environments, it is unclear how they are formed. The proposed formation mechanisms are: i) they are the remnant nuclei of dwarf galaxies that have been tidally stripped \citep{bekki2001galaxy, bekki2003galaxy, drinkwater2003class, pfeffer2013ultra}; ii) they are the upper tail of the mass distribution of globular clusters \citep{hilker1999central, mieske2004fornax, evstigneeva2007internal, forbes2008uniting}; or iii) the merging of star clusters to form very massive ‘super’ clusters \citep{kroupa1998dynamical, fellhauer2002formation, bruns2011parametric}. It has been suggested that the high mass-to-light ratios characteristic of UCDs may be the result of top-heavy stellar IMF \citep{dabringhausen2009top}, and that UCDs with top-heavy IMFs may appear as quasar-like objects \citep{jerabkova2017formation}.

The properties of UCDs are similar to those of nuclear clusters: as well as those listed above, the metallicities \citep{brodie2011relationships, francis2012chemical}, internal velocity dispersions \citep{drinkwater2003class} and colour-magnitude relationships \citep{cote2006acs,evstigneeva2008structural, bruns2011parametric} are similar to those of dwarf galaxy nuclei. Simulations of the threshing scenario performed by \cite{bekki2001galaxy,bekki2003galaxy} produce objects with sizes and luminosities comparable with observed UCDs. However, the simulations from \cite{pfeffer2014contribution} show that tidal threshing accounts for only approximately half of observed UCDs, and tidal threshing does not necessarily reproduce some low-mass extended UCDs observed in the Fornax \citep{thomas2008formation} and Virgo clusters \citep{brodie2011relationships}. 

Though there has been extensive investigation into how UCDs form by numerical simulation, so far the problem has been investigated primarily through collision-less simulations in which massive star cluster systems are pre-existing \citep{pfeffer2014contribution}. These studies have been conducted under the assumption that the galaxies have a high fraction of nucleation. Massive nuclear star clusters would most likely form from giant molecular clouds (GMCs) and migrate to the centre as a result of dynamical friction. This process in dwarf galaxies has been investigated using numerical simulation \citep{Guillard2016}, however, it has not been investigated at very high resolutions and with specific reference to UCD formation. A thorough understanding of this process is necessary to make detailed predictions about key observables in UCDs. 

The aim of this work is to numerically investigate the mechanism by which massive nuclear star clusters might form, before undergoing the tidal threshing processes already investigated. Using a dynamical model of giant molecular clouds, the impact of feedback effects of asymptotic giant branch (AGB) winds and first-generation stars becoming Type-II supernovae (SNII) on the formation of high density nuclei within dwarf galaxies is investigated. We will discuss whether the simulated UCDs have scaling relations that are consistent with observations. 

The plan of the paper is as follows: the first section is an introduction to the topic of UCDs. Section 2 will detail the model used to investigate the dynamical evolution of massive nuclear star clusters from GMCs. Section 3 presents the results of the star clusters formed from GMCs with a focus on the development of dense nuclei. Comparison of these results with those from recent observational and theoretical studies will be in Section 4, and a summary of any conclusions will be given in Section 5.


\begin{table}
\centering
\begin{minipage}{85mm}
\caption{Description of key physical properties  for
the fiducial UCD formation model.}
\begin{tabular}{ll}
{ Parameters } &
{ Values } \\
Initial SGMC mass   & $10^8 {\rm M}_{\odot}$  \\
Initial SGMC size   & $200 $pc  \\
Fractal dimension   & $D_3=2$  \\
Power-law slope    & $\beta=1$  \\
Initial virial ratio   & $t_{\rm vir}=0.34$  \\
Rotational energy fraction   & $t_{\rm rot}=0$  \\
Initial number of gas particles  & 1048911 \\
Mass resolution  &  $9.5 \times 10 {\rm M}_{\odot}$ \\
Size resolution  &  0.39 pc \\
SN and AGB feedback    & Yes \\
AGB yield  & \cite{karakas2010updated} \\
SN yield  & T95 \\
Threshold gas density for star formation  & $10^4$ cm$^{-3}$ \\
\end{tabular}
\end{minipage}
\end{table}

\section{The model}

\subsection{A SGMC scenario}

\begin{figure*}
	\includegraphics[width=\linewidth]{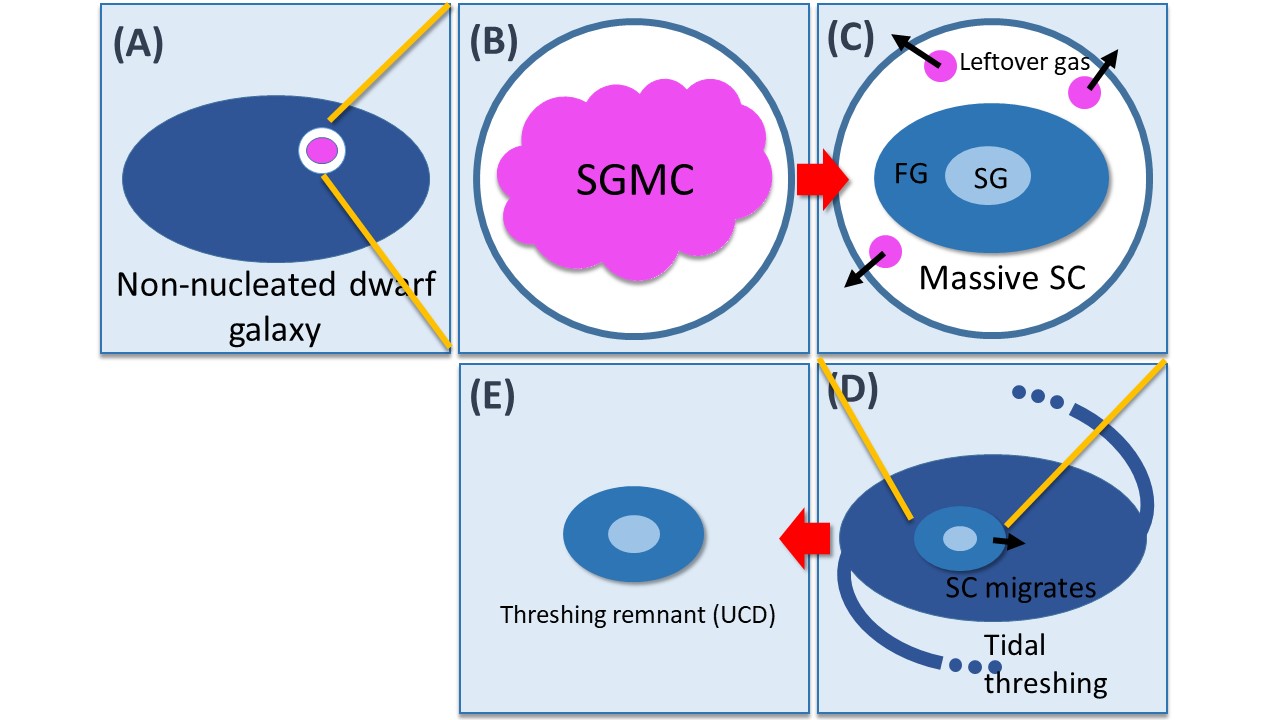}
	\caption{An overview of the investigated scenario for UCD formation. Four distinct and important phases of the UCD evolution scenario are shown in panels (\textbf{A})-(\textbf{E}), in chronological order. After the formation of a non-nucleated dwarf galaxy (\textbf{A}), an SGMC forms somewhere within the dwarf galaxy (\textbf{B}). This SGMC collapses, initiating star formation, resulting in a massive star cluster with distinct populations (\textbf{C}). The massive SC migrates to the centre of the dwarf galaxy, which is tidally stripped or 'threshed' by the host cluster (\textbf{D}) and the remnant, primarily composed of the massive SC, is now observable as a UCD (\textbf{E}).}
	\label{fig:scenario}
\end{figure*}

Here we investigate the following formation route for UCDs, a graphic representation of which is given in Fig. \ref{fig:scenario}. Some time after the formation of a non-nucleated dwarf galaxy, a super giant molecular cloud (SGMC) of mass $10^7 - 10^8 {\rm M}_{\odot}$ forms. The collapse of this SGMC initiates violent starbursts, producing a stellar population. After some fraction of this population become AGB stars, they will expel sufficient gas to accumulate and begin a second round of star formation. This produces a second stellar population and forms a massive star cluster within the dwarf galaxy. This process may repeat several times as earlier populations age, producing multiple stellar generations. Over time, this massive SC will migrate to the central region of the dwarf galaxy as a result of dynamical friction. This massive nuclear star cluster may then have the outer envelope of stars stripped from it, or 'threshed', by the strong tidal gravitational field of a galaxy cluster. The remnant of this threshing process is then identifiable as a UCD. Star formation processes during UCD formation have so far not been investigated in previous numerical simulations of UCD formation, and the  SF histories of UCDs will be able to provide valuable clues as to the photometric properties of UCDs at their formation, as discussed recently in \cite{jerabkova2017formation}. It should be noted here that \cite{harris1994supergiant} pointed out the importance of SGMCs in GC formation for the cases of very low star formation efficiencies of GMCs. This scenario required even larger masses of GMCs in order to explain the observed large masses of stars in UCDs.

GMCs in the Galaxy and nearby galaxies are observed to have fractal structures (see \cite{blitz2006giant}), so for the simulations in this study we have adopted fractal structures in SGMCs. The fractal dimension ($D_3$) is an important parameter of fractal GMCs in three-dimensional space, and is derived observationally for the interstellar medium (ISM) of the LMC and the SMC (e.g. \cite{Sun2017}). We choose a value of $D_3=2$ for all models in this study as it is more consistent with $\beta = 1$ \citep{Bekki2017a} but the fractal dimension is not as important in this study as other initial parameters of SGMCs.

We have run smooth particle hydrodynamics (SPH) simulations of UCD formation by adapting our own original simulation code that can be run on GPU clusters \citep{bekki2013galactic, bekki2015formation,Bekki2017a}. The details of the code for GC-scale simulations are given
in \cite{Bekki2017a} (hereafter B17) and we consider that code appropriate for use in the investigation of UCD formation. The code includes various physical processes of the ISM, such as the formation of molecular hydrogen (${\rm H_2}$) from neutral H on dust grains, dust formation,  destruction, and growth, effects of photo-electric heating on cold gas, star formation, SN feedback effects on star formation, which we did not include in the present simulation of UCD formation. This is mainly because our main focus in not UCD formation regulated dust. The initial smoothing length is set to be the same as the gravitational softening length, but it changes with time according to the density of gas particle,  i.e.,  adaptive smoothing length, which is the standard method in the SPH algorithm. Radiative feedback effects in SGMCs have been ignored for the present investigation. Several other environmental effects that are important in cluster formation from GMCs have here been ignored due to the mass and density of the SGMCs: \cite{Dale2012} shows that ionizing feedback has little effect on GMCs of mass greater than $10^{6} \textrm{M}_{\odot}$, and so will not apply to our scenario. SN feedback effects are important for lower-mass GMCs forming GCs but these effects become negligible at the scales we are investigating here \citep{Ashman2001}. Radiation pressure from developing SCs within GMCs has been shown by \cite{Murray2009} to have some effect on star formation but the importance of this effect scales with the mass and density of the GMC. The mass of the fiducial model ($M_{SGMC}= 10^{8} \textrm{M}_{\odot}$) is lower than the mass of BX482 simulated by \cite{Murray2009}, which has SGMC mass $M_{SGMC}= 10^{9} \textrm{M}_{\odot}$, but over a far greater volume than our fiducial model.

\subsection{SGMC}

A SGMC with mass of $M_{\rm mc}$ and a size of $R_{\rm mc}$ is assumed to have a power-law radial density profile ($\rho_{\rm mc}(r)$):
\begin{equation}
\rho_{\rm mc} (r)=\frac{\rho_{\rm mc,0}}{ (r+c_{\rm mc})^{\beta} },
\end{equation}
where $r$,  $\rho_{\rm mc, 0}$, and $c_{\rm mc}$, $\beta$ are the distance from the centre of the SGMC, a constant determined by $M_{\rm mc}$ and $R_{\rm mc}$, the radius of the core of the MC, and the slope of the power-law. In all equations we have used "mc" in place of "SGMC" for simplicity. We adopted $\beta=1$ for the profiles and investigated SF histories from gas at UCD formation. B17 adopt the following mass-size relation of GMCs:
\begin{equation}
R_{\rm mc}
 =\rm C_{\rm mc} \times  (\frac{M_{\rm mc}}{5 \times 10^5  {\rm M}_{\odot} })^{0.53}
{\rm pc}
\end{equation}
In this study we have chosen a value of $\rm C_{\rm mc} = 40$, consistent with observations (B17). The choice of $\beta=1$ for the power-law slope is consistent with this mass-size relation, as the total mass of a GMC is approximately proportional to $R^{3-\beta}$ or here $M_{\rm mc} \propto R_{\rm mc}^{2}$. It is not clear whether progenitor GMCs of UCDs will follow the same relation. Therefore, we investigate different $R_{\rm mc}$ for a given $M_{\rm mc}$. A more complete description of how the initial fractal distribution can be set up is given in B17.

As shown in B17, the initial virial ratio ($t_{\rm vir}$) can determine the total amount of kinetic energy ($T_{\rm kin}$) of a SGMC, which can be important for the formation of UCDs. The ratio is described as $t_{\rm vir}=\frac{ 2T_{\rm kin} }{ | W_{\rm mc} | }$, where $W_{\rm mc}$ is the initial total potential energy of the SGMC. The models with $t_{\rm vir}=0.35$ are mostly investigated in the present study, because, as shown in B17, the simulated stellar systems can be compact like GCs and UCDs for that $t_{\rm vir}$. Following B17, we consider that (i) a SGMC can have rigid rotation and thus (ii)  $T_{\rm kin}$ is the combination of the total random energy $T_{\rm ran}$ and the total rotational one ($T_{\rm rot}$). We accordingly introduce the parameter $f_{\rm rot}=\frac{ T_{\rm rot} }{T_{\rm kin}}$ which is set to be either 0 (no rotation) or 0.1 (with rigid rotation). The way to implement this global rigid rotation in initial SGMCs is given in B17.

In these simulations, each SGMC is represented by $N \approx 10^6$ gas particles. This is enough to investigate star formation processes on the parsec scale (i.e., not the formation of individual star) and chemical enrichment via SNII and AGB stars feedback (B17). For the mass resolution ($9.5\times10 \textrm{M}_{\odot}$) used in this investigation, stochastic sampling of the stellar IMF may influence the results, if massive stars are preferentially selected due to even stronger SNII feedback effects. Initial gaseous temperature and metallicity are set to be 10K and [Fe/H]=$-1.6$ in all MCs; variations in initial [Fe/H] show no impact on star formation in the fiducial model and so initial metallicity values are not changed in any other models. The radiative cooling processes are included, using the cooling curve by \cite{rosen1995global} for  $T < 10^4$K and the MAPPING III code for $T \ge 10^4$K \citep{sutherland1993cooling}. Cooling by ${\rm H_2}$ gas and heating by dust and other processes are not included in the present study.

\subsection{Star formation, SN feedback, and chemical enrichment}

Here we will cover in brief the details of the models for star formation, SN feedback effects on gas and chemical enrichment from SNeII and AGB stars, which are described in full in B17. New stars can be formed from gas particles if the following two physical conditions are met. The first is that the local density ($\rho_{\rm g}$) exceeds a threshold (${\rho}_{\rm th}$) of star formation ${\rho}_{\rm g} >{\rho}_{\rm th}$.
In the present study only one new star is generated from each SPH gas particle (i.e., not multiple times) in order to avoid a dramatic increase of particle number during a simulation. It is assumed that star formation will occur in the dense cores of MCs. Subsequently, $\rho_{\rm th}$ is set to be $[10^4-10^5]$ H atoms cm$^{-3}$, which is consistent with the observed values \citep{bergin2007cold}. The local velocity field around a gas particle is consistent with that for gravitationally collapsing, and is $div {\bf v}<0$.

New star particles are created with a fixed IMF and an initial mass $m_{\rm ns}$. Mass loss by SNe Ia, SNe II and AGB stars reduces stellar mass over time, and the final stellar mass at $T=3\times10^{8}$ years (the duration of the simulation) may differ significantly from  $m_{\rm ns}$. The IMF we have adopted is defined as $\psi (m_{\rm s}) = C_i{m_{\rm s}}^{-\alpha}$, where $m_{\rm s}$ is the initial mass of each individual star and the slope $\alpha =2.35$ corresponds to the Salpeter IMF \citep{Salpeter1955}. The Salpeter IMF has been chosen for these models to facilitate simple calculation of total AGB mass ejecta, the number and mass of SNII present analytically within the simulation. The normalization factor $C_i$ is a function of stellar particle mass, $m_{\rm l}$ (lower-mass cutoff), and $m_{\rm u}$ (upper-mass cutoff):
\begin{equation}
C_i=\frac{m_{ns} \times (2-\alpha)}{{m_{\rm u}}^{2-\alpha}-{m_{\rm l}}^{2-\alpha}}.
\end{equation}
where $m_{\rm l}$ and $m_{\rm u}$ are set to be $0.1 {\rm M}_{\odot}$ and $120 {\rm M}_{\odot}$, respectively. We have investigated only the models with $\alpha=2.35$ in the present study.

SNe of new stars can thermally and kinematically perturb the surrounding gas within GC-forming MCs. Each SN is assumed to produce the feedback energy ($E_{\rm sn}$) of $10^{51}$ erg which is then converted into kinematic and thermal energy of the gas surrounding the SN. \cite{thornton1998energy} investigated the fraction of $E_{\rm sn}$ produced by a SN that could create random motion in local gas ("kinematic feedback"). In this simulation, multiple SNe can occur within one MC at different epochs $\sim 30$ Myr after star formation, and as a result the ratio of kinematic feedback to total SN energy ($f_{\rm kin}$) can be different to those predicted by previous simulations of single SN. The interaction of expanding shells of different SNe leads $f_{\rm kin}$ to be quite high. The details of how  $f_{\rm kin}E_{\rm sn}$ (i.e., kinetic feedback energy) of SNe is distributed amongst neighbouring gas particles are provided in B17. The time delay between conversion of gas into a new star and a SN occurring is parametrized by $t_{\rm delay}$, which depends on the initial stellar masses.

SN gas ejecta will mix with local gas particles, resulting in an increase of their chemical abundances. The chemical abundance of $k$th element ($k$=1, 2, 3, ... correspond to H, He, C, N, O. ... respectively) for $j$th gas particle ($Z_{j, k}$) among $N_{\rm sn}$ surrounding gas particles around a SN can change according to the following equation:
\begin{equation}
(m_{j}+m_{\rm ej}) Z_{j,k}^{'} = m_{j}Z_{j,k}+
\frac{ \Delta m_{\rm ej} Z_{\rm sn, \it k} }{ N_{\rm sn} },
\end{equation}
where $Z_{j,k}^{'}$ are the chemical abundance of $k$th element after chemical enrichment by the SN and $Z_{\rm sn, \it k}$ is the chemical abundance of $k$th element for the SN ejecta. Here the chemical yield table of SNII from \cite{tsujimoto1995relative} (hereafter T95) has been used to calculate $Z_{\rm sn,k}$.

\subsection{Gas ejection and feedback effects from AGB stars}

Since we adopt the same models as those used in B17 for the present investigation of UCD formation, we describe the models briefly here. We consider that (i) AGB stars can eject gas significantly later than SNe and therefore (ii) SN explosion can expel almost all of original cold gas around
intermediate-mass stars. A novel model (B17) in which each AGB star eject gas particles with chemical abundances predicted from recent AGB models (e.g., K10) is adopted so that chemical pollution by AGB stars can be implemented in hydrodynamical simulations of UCD formation. Ejection of new particles from AGB stars (`AGB particle') means that the total number of gas particles can significantly increase as a simulation goes. In order to model chemical enrichment by AGB stars with different masses more properly, we consider ejection of AGB particles at different epochs depending the masses of AGB stars (See B17 for the details.)

An `AGB gas particle' is ejected from a new stellar particle with a wind velocity of $v_{\rm wind}$
at the end of the  main-sequence phase of the stellar particle. Although this $v_{\rm wind}$ is an order of 10 km s$^{-1}$, such stellar wind can dramatically influence the star formation histories within existing SCs (e.g., \cite{DErcole2010}, \cite{Bekki2010}, \cite{Bekki2011}, and B17) We adopt $v_{\rm wind}= 10$  km s$^{-1}$, which is
consistent with recent observations of AGB stars in the LMC (e.g., \cite{marshall2004asymptotic}). The initial temperature of AGB wind ($T_{\rm wind}$)  is set to be 1000 K, which is consistent with standard theoretical  models of  AGB winds. It is likely that SG formation from gas is possible only
if stellar wind can be efficiently cooled down from $T_{\rm wind}$ to a few tens K.

\subsection{Parameter study}

We consider that  $M_{\rm mc}$ and  $R_{\rm mc}$ are the key parameters in the formation processes of UCDs in the present study, because other possibly important parameters such as IMF slope are fixed.
We mainly describe the results of the fiducial model in which
$M_{\rm mc}=10^8 {\rm M}_{\odot}$,
$R_{\rm mc}=200$ pc,
$f_{\rm rot}=0$, and
$D_{\rm 3}=2$,
because this model shows the typical behavior of UCD formation within fractal MCs. The basic parameters used for the fiducial model are summarized in Table 1. We also discuss the results of other models with different values of the key parameters. The parameter values of all 15 models discussed in this paper are summarized in Table 2. The model M1 was run as a low resolution test of the simulation code, the results of which were sufficient to warrant further, higher resolution models; however, the low resolution model is insufficient for our purposes here. Hence, M2 is the fiducial model in this study.

\begin{table}
\centering
\begin{minipage}{85mm}
\caption{The basic model parameters for
fractal supergiant molecular clouds (SGMCs).}
\begin{tabular}{lllll}
{ Model ID } &
{ $M_{\rm mc}$ \footnote{ The initial total mass of
a fractal molecular cloud (MC) in units of $10^7 {\rm M}_{\odot}$.
 }} &
{ $R_{\rm mc}$ \footnote{ The initial size for a MC
in units of pc.
 }} &
{ $f_{\rm rot}$  \footnote{ The initial ratio of total rotational
energy to total kinetic energy in a MC.
}} & Model notes\\
M1 & 10 & 200 & 0.0 & low res. test model\\
M2 & 10 & 200 & 0.0 & \\
M3 & 10 & 200 & 0.1 & \\
M4 & 3 & 100 & 0.0 & \\
M5 & 3 & 100 & 0.1 & \\
M6 & 10 & 200 & 0.0 & SNeII, no AGB wind\\
M7 & 10 & 200 & 0.0 & SNeII and AGB wind \\
M8 & 10 & 200 & 0.0 & no SNeII/AGB wind\\
M9 & 10 & 200 & 0.0 &${\rho}_{\rm th}=10^{5} \textrm{cm}^{-3}$\\
M10 & 0.1 & 20 & 0.0 & \\
M11 & 1 & 63 & 0.0 & \\
M12 & 3 & 35 & 0.0 & \\
M13 & 3 & 57 & 0.0 & \\
M14 & 5 & 45 & 0.0 & \\
M15 & 5 & 89 & 0.0 & \\
\end{tabular}
\end{minipage}
\end{table}

\section{Results}

\subsection{Fiducial model}

\begin{figure*}
	\includegraphics[width=\linewidth]{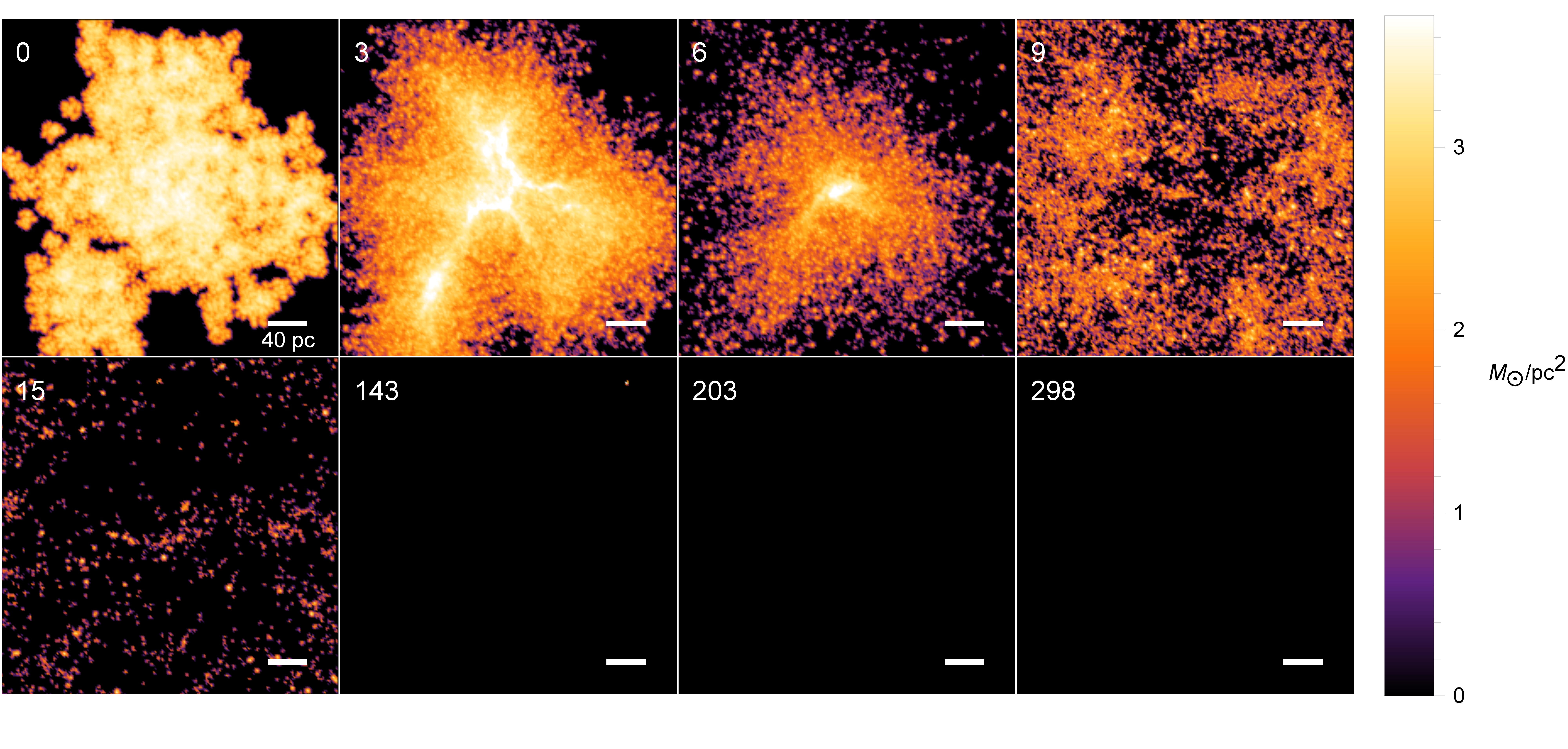}
	\caption{Evolution of projected mass density of initial SGMC gas in the fiducial model M2 in the xy-plane. The time T in Myr indicated in the upper left corner of each frame represents the time elapsed since the beginning of the simulation. The frames are 200pc by 200pc. Density is given in units of $M_{\odot}/pc^{2}$ as indicated by the legend to the right.}
	\label{fig:ucd2FGgas}
\end{figure*}

\begin{figure*}
	\includegraphics[width=\linewidth]{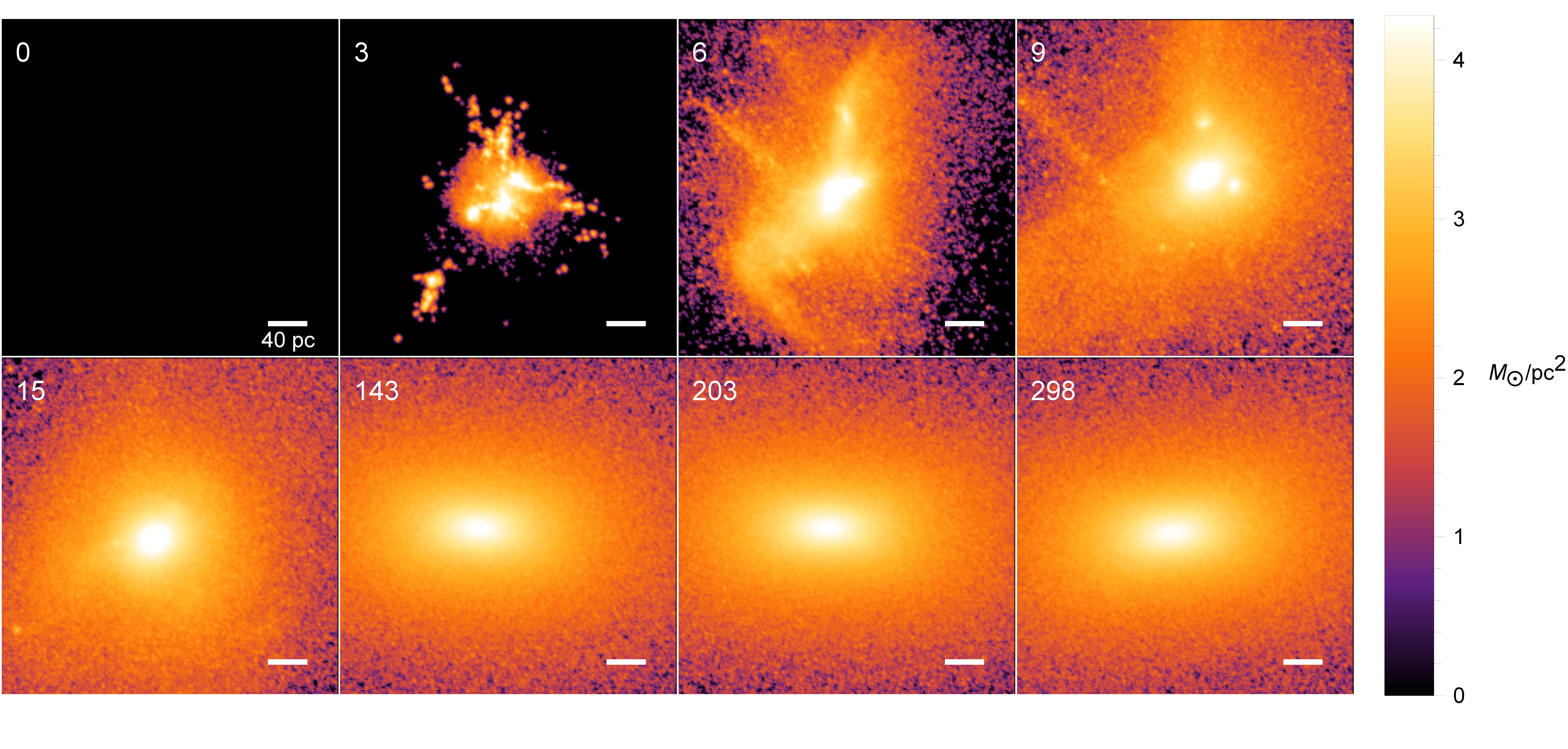}
	\caption{Projected mass density of first-generation stellar components of the fiducial model M2, shown in the xy-plane. The time T in Myr indicated in the upper left corner of each frame represents the time elapsed since the beginning of the simulation. The frames are 200pc by 200pc. Density is given in units of $M_{\odot}/pc^{2}$, indicated by the legend to the right.}
	\label{fig:ucd2FGstars}
\end{figure*}

Fig. \ref{fig:ucd2FGgas} and Fig. \ref{fig:ucd2FGstars} show the two-dimensional projected mass density of the initial gas particles and the first generation (FG) stars in the fiducial model M2, respectively. As the SGMC collapses, pockets of density greater than ${\rho}_{\rm th}$ rapidly form, within $T < 3$ Myr, where T is the time elapsed since the beginning of the simulation. In these pockets there is an initial, intense burst of star formation, forming small star clusters. Within $T\approx30$ Myr the majority of the SGMC gas has been consumed in rapid star formation and any remaining gas is expelled over time by SNII feedback; this dispersion is clearly visible at T = 9 Myr and T = 15 Myr. During this period, the smaller clusters of FG stars fall inwards and merge, and by $T\approx50$ Myr the first-generation stars have stabilised into an elliptical distribution with a half-mass radius of $R\approx60$ pc. This distribution remains consistent throughout the rest of the simulation.

\begin{figure*}
	\includegraphics[width=\linewidth]{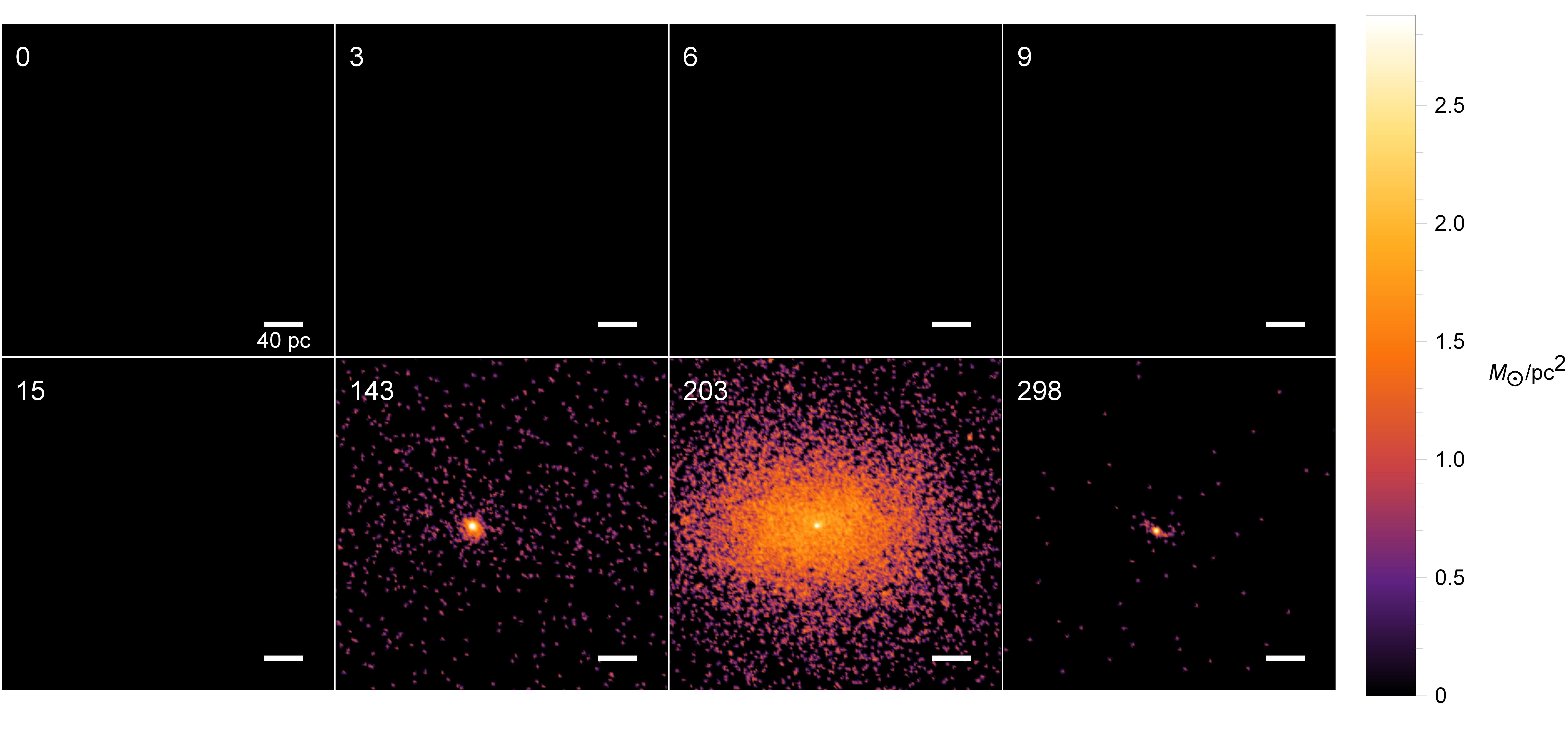}
	\caption{Projected mass density of AGB gas ejecta in the fiducial model M2.}
	\label{fig:ucd2AGBgas}
\end{figure*}

\begin{figure*}
	\includegraphics[width=\linewidth]{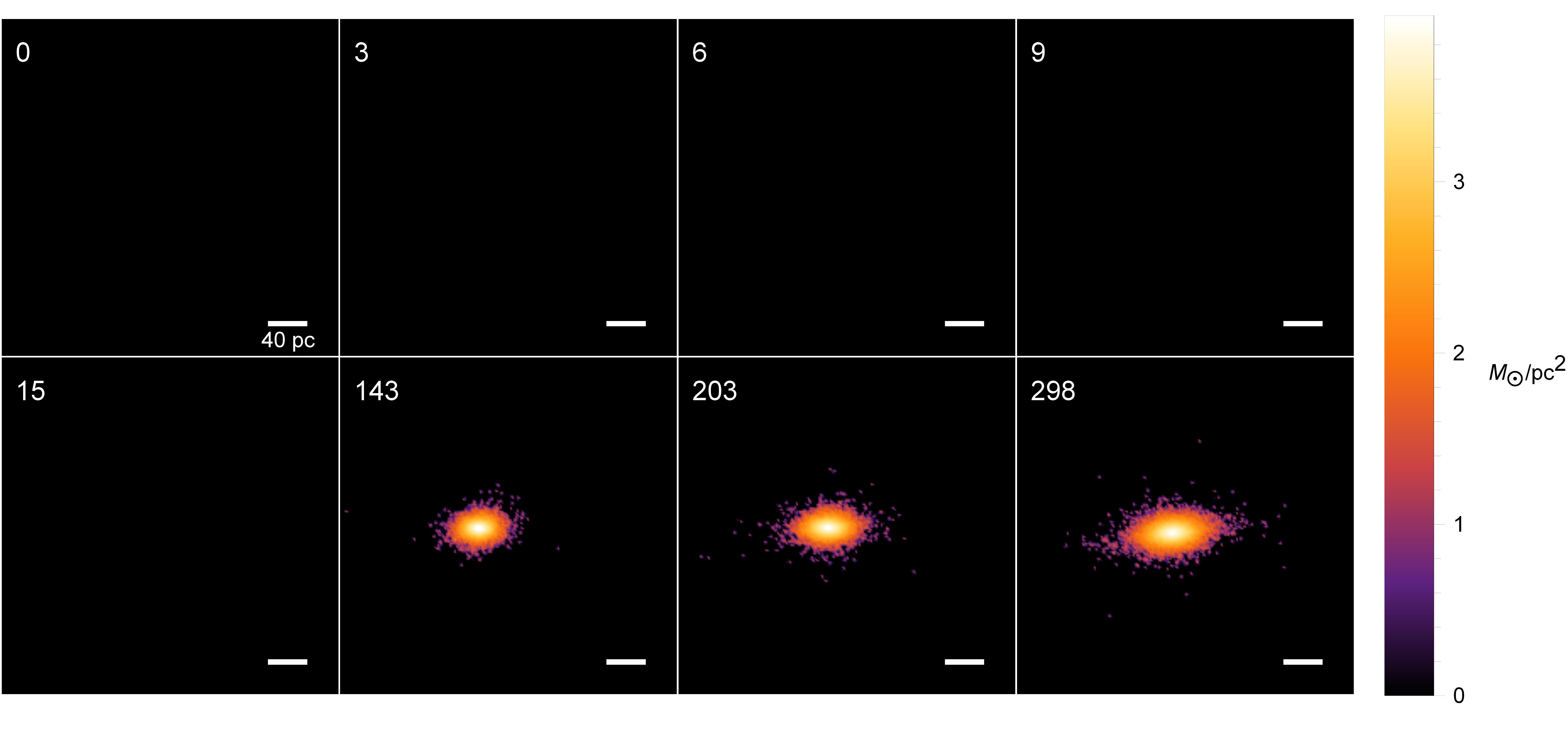}
	\caption{Projected mass density of SG (those formed from AGB gas) stellar components of the fiducial model M2. }
	\label{fig:ucd2AGBstars}
\end{figure*}

\begin{figure}
	\includegraphics[width=\linewidth]{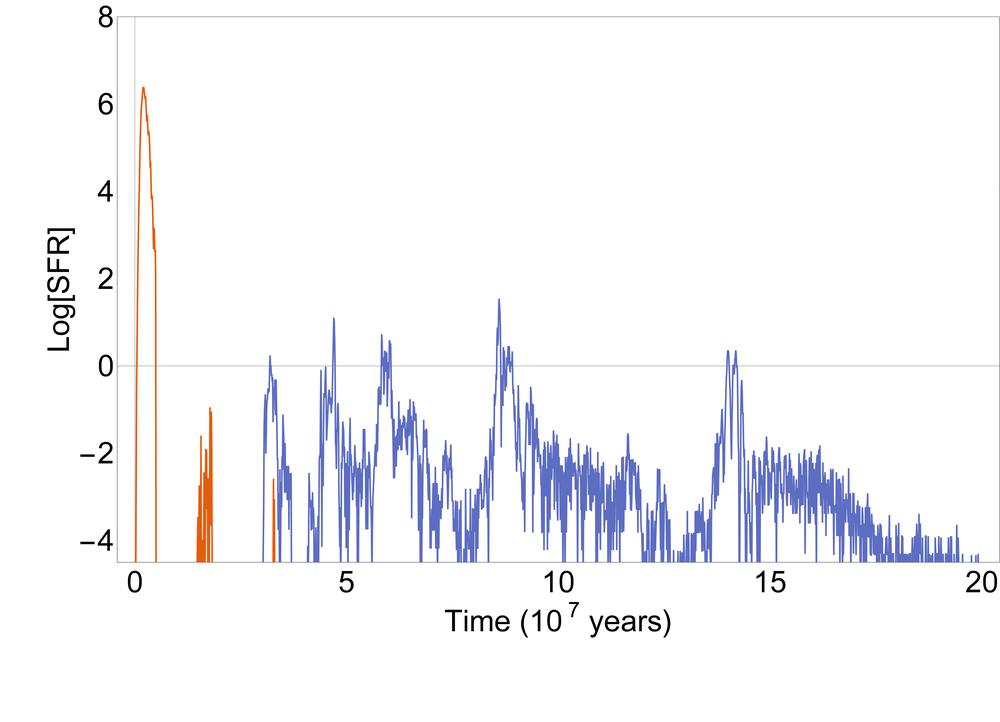}
	\caption{Logarithm of the star formation rates in the first and second generations of simulation 2. The orange line denotes star formation in the FG, the blue line star formation in the SG.}
	\label{fig:ucd2sfr}
\end{figure}

\begin{figure}
	\includegraphics[width=\linewidth]{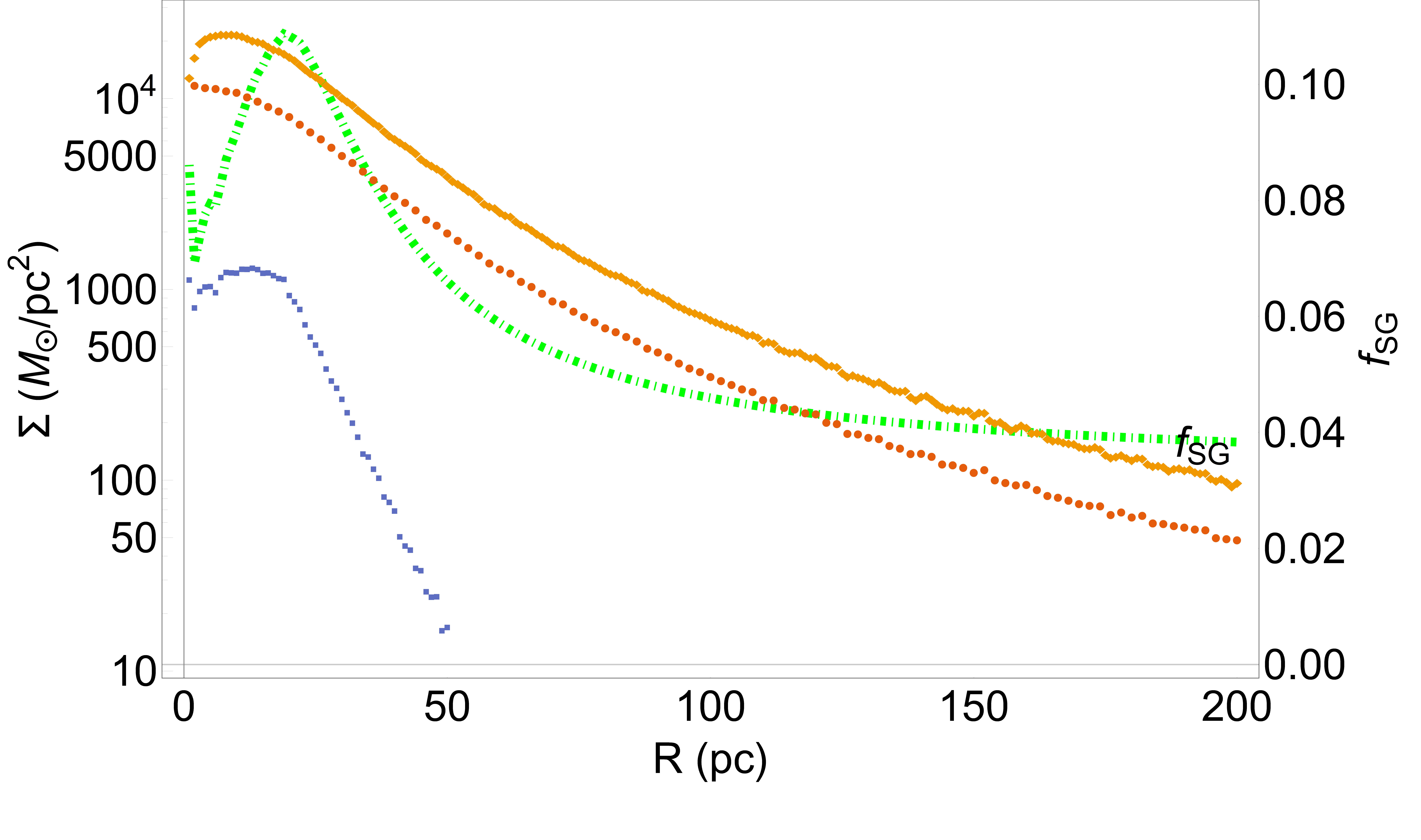}
	\caption{Surface mass density profiles of FG stars (in orange), SG stars (in blue) and of both components (in yellow) in M2 at the final timestep ($T=298$ Myr).  The centre of mass of the SG is clearly misaligned with that of the FG. This is emphasised in the plot of $ \rm f_{\rm SG}=\frac{\rm M_{\rm SG}}{\rm M_{\rm FG}+\rm M_{\rm SG}}$ shown as a green dot-dashed line. The scale of $\rm f_{\rm SG}$ is given on the right hand y-axis.}
	\label{fig:ucd2SB}
\end{figure}

Fig. \ref{fig:ucd2AGBgas} shows the projected mass density of gas particles ejected from FG AGB stars and Fig. \ref{fig:ucd2AGBstars} that of second generation (SG) stars. FG AGB begin to eject gas particles at a significant enough rate to start a second round of star formation by $T\approx30$ Myr, as shown in Fig. \ref{fig:ucd2sfr}. This gas is emitted in and remains primarily within the central region of the cluster and so by $T=143$ Myr there is a dense cluster of SG stars with a half-mass radius of $R \approx 10$ pc. Further star formation does not significantly increase the size of this cluster, only the density. Large fluctuations in the quantity of AGB gas present in the system are evident in the last three timesteps of \ref{fig:ucd2AGBgas}. AGB gas accumulates in the central regions until there is sufficient density to cause another peak in star formation; any outlying remainder is then likely expelled by stellar winds, and the cycle repeats. By $T=298$ Myr two possible proto-arms are clearly visible, but the model does not exhibit significant rotation overall. The nested structure of the SG stars is visible in the radial mass density profile shown in Fig. \ref{fig:ucd2SB}: the SG stars are strongly constrained within $R=50$ pc, with a half-mass radius of $\rm R_{\rm h}=18$ pc, while the FG of stars extends out to $R=200$ pc. The centre of mass of the SG is visibly distinct from that of the FG, and this could give rise to the core-like central distribution. The fraction of SG stellar mass is given:
\begin{equation}
\rm f_{\rm SG} = \frac{\rm M_{\rm SG}}{\rm M_{\rm FG}+\rm M_{\rm SG}}
\end{equation} 
This reaches a maximum value of $\rm f_{\rm SG} \approx 0.1$ at $\rm R \approx 20$ pc, noticeably misaligned with the centre of mass of the FG. While the system is not perfectly spherically symmetrical, there is no noticeable difference between the centre of potential and the centre of mass in these models, and the centre of mass has been subsequently used for all analysis.  $\rm f_{\rm SG}$ declines sharply beyond this peak, further illustrating the compact nature of the SG population. SG stars, having formed from AGB gas, are more He-rich than the FG, and subsequently can appear bluer, as in \cite{norris2004helium}, \cite{DAntona2002} and \cite{D’Antona2004}. Then $\rm f_{\rm SG}$ is an indication of the colour gradient of the cluster, showing a bluer core and a redder envelope of FG stars. There is no noticeable variation in the metallicity of the overall SC from the initial value of [Fe/H] = -1.6.

Fig. \ref{fig:ucd2sfr} shows the star formation rates of FG and AGB stars in M2: most FG stars are formed within the first $10$ Myr in the initial star formation burst. Ongoing star formation is not present; SN feedback effects act during this time to expel any remaining SGMC gas. At around 30 Myrs gas lost by AGB stars has accumulated in sufficient density to start SG star formation. As noted above, SG star formation occurs in 'waves' as more of the original stars and earlier SG stars begin to expel gas into the system. Peaks in the star formation rate are visible at around $T=50, 90$ and $140$ Myrs into the simulation, but some amount of SG star formation is ongoing until $T=200$ Myr after the start of the simulation, at which point the AGB gas within the system is exhausted and star formation ceases. 

\begin{figure}
	\includegraphics[width=\linewidth]{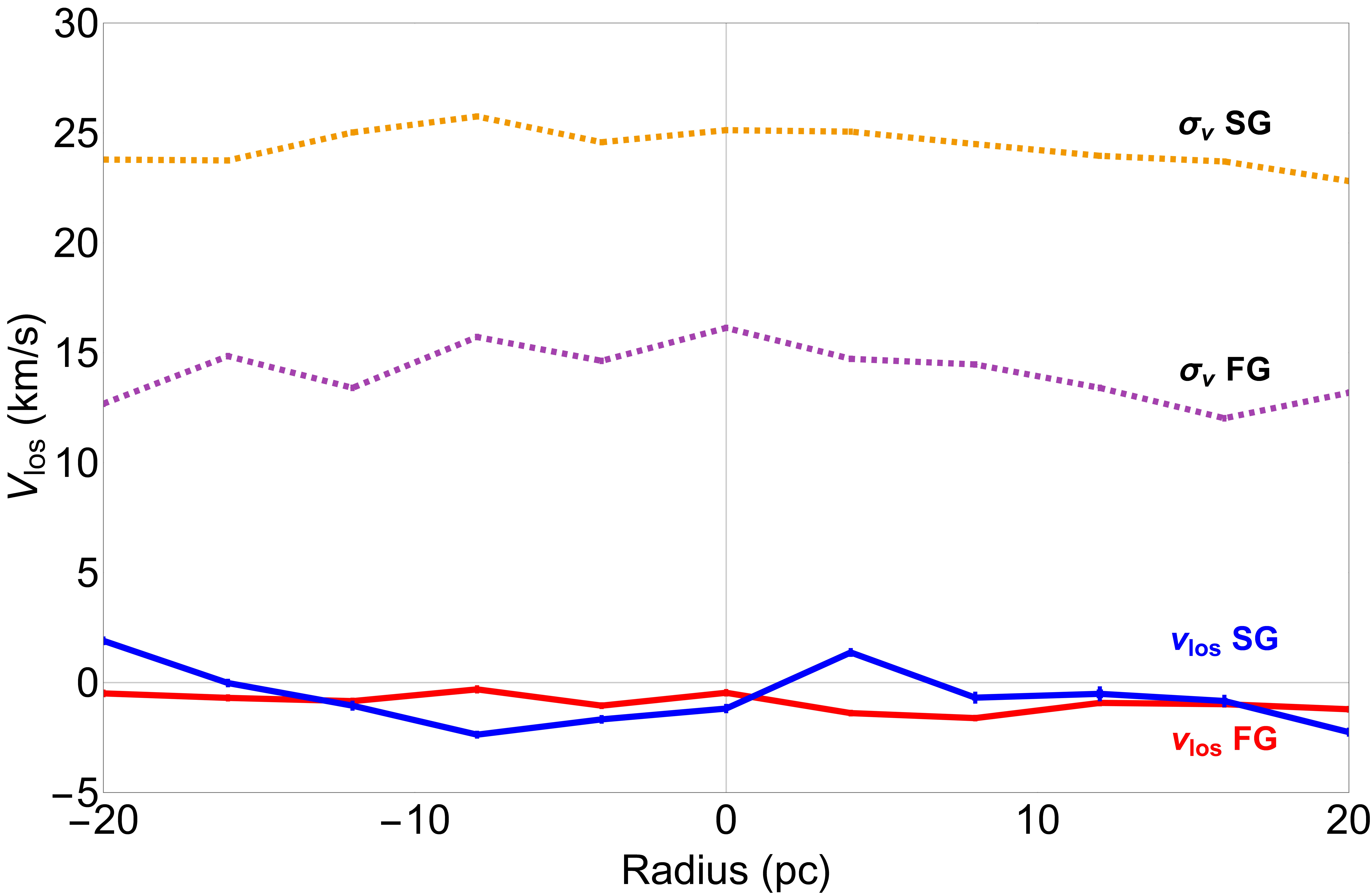}
	\caption{Line of sight velocity of first (solid red) and second (solid blue) generation star populations and $\sigma_{v}$ of first (dashed yellow) and second (dashed purple) generation star populations in M2.}
	\label{fig:ucd2vlos}
\end{figure}

\begin{figure*}
	\includegraphics[width=\linewidth]{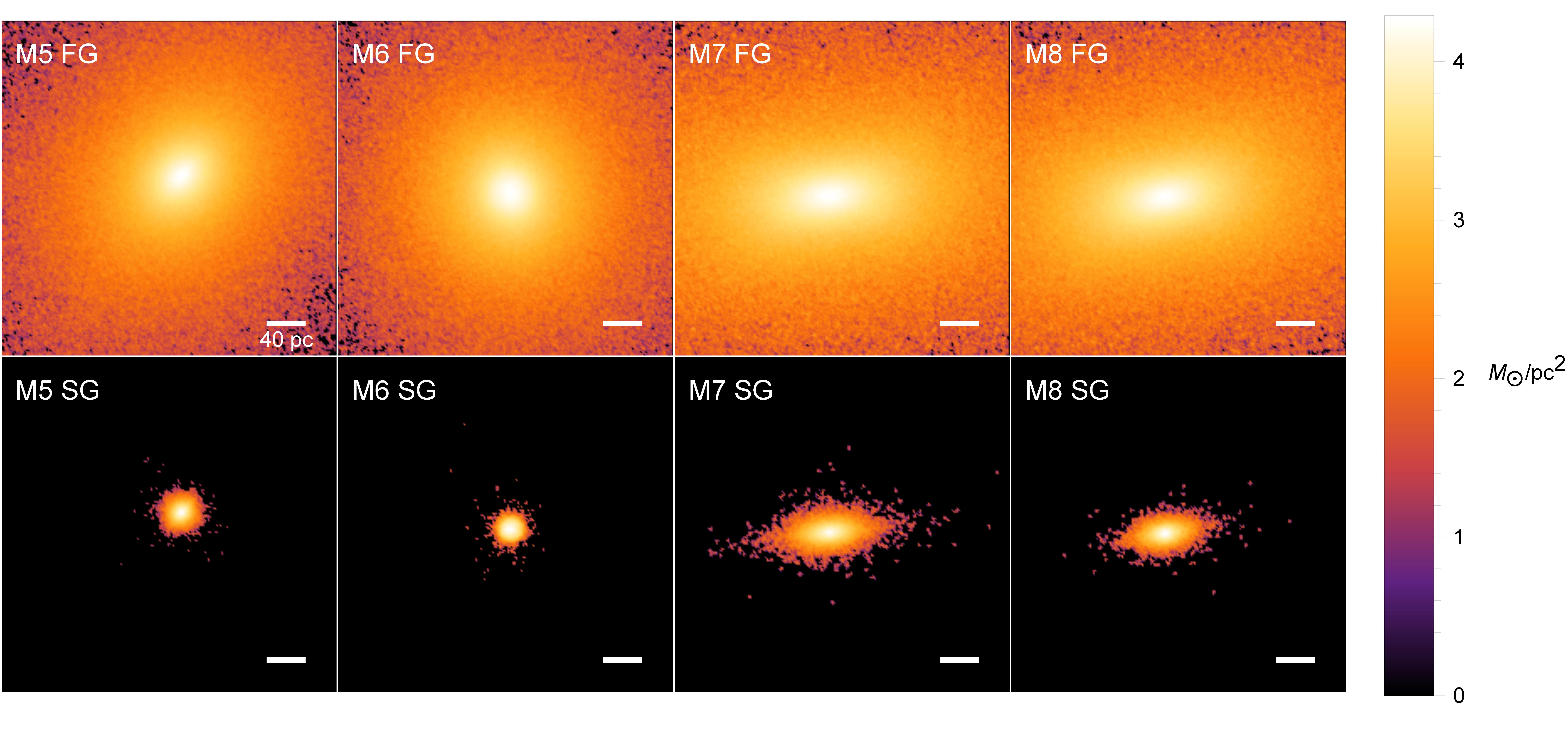}
	\caption{Projected mass density of FG and SG stars in the xy-plane of four secondary models, M4, M5, M6 and M7, from left to right. The top panels show the projected mass density of the FG stars during the final timestep $T=298$ Myr, and the bottom row shows the projected mass density of the SG stars during the final timestep.}
	\label{fig:ucds}
\end{figure*}

The FG stars exhibit little to no overall rotation, as shown in Fig. \ref{fig:ucd2vlos}, as the SGMC gas from which they formed was non-rotating. However, the highly clustered SG exhibit some rotation within $R<10$ pc of up to $v_{r}\approx2.5$ km s$^{-1}$. Both populations have low velocity dispersions, peaking at $\sigma_{v}\approx26$ km s$^{-1}$ for FG and $\sigma_{v}\approx16$ km s$^{-1}$. There are what appear to be central gradients in the $\rm v_{\rm los}$ plots of SG stellar populations (see Fig. \ref{fig:ucd2vlos}) but these are not significant enough for us to suggest confidently that there is rotation the in SG population. Nor do those models showing significant quantities of AGB gas remaining at the final timestep display significant rotation of this gas. 

\subsection{Parameter dependence}

The model M3 begins with a GMC of the same mass and dimensions as the main model M2, however, the cloud has initial rigid rotation of $f_{\rm rot}=0.1$. While star formation from the GMC gas occurs, no SG stars or AGB gas is present at the end of the simulation. The rotation is clearly visible, with radial velocities of $v_{\rm rot}\approx20$ km/s amongst the stars within $R<10$ pc, and this rotation likely causes any AGB gas to be rapidly expelled from the system before any SG stars can form. Unlike M2, there is some SGMC gas remaining by the final timestep $T=298$ Myr, and by this time it has started to spiral back into the central regions of the system after wider dispersion.

Fig. \ref{fig:ucds} shows the distributions of FG and SG stars in the models M4, M5, M6 and M7 at $T=298$ Myr. The model M5 is more compact and less massive than M2, with no initial rotation. This variation still results in the two generation nested structure visible in M2, with a less extended halo of FG stars ($r_{\rm h}=49$ pc). SGMC gas has completely dissipated, but AGB gas is still dense in the central parts of the star cluster, potentially indicating further SG star formation. This is similar to M5, which has the same mass and dimensions as M4 but is initially rotating, leading to wider distribution of FG stars. Star formation in M5 is slowed by the rotation of the system, resulting in fewer SG stars by $T=298$ Myr, but a greater amount of AGB gas present in sufficient density for ongoing star formation within the central region. The two generation nested structure is also present in M6 and M7, both of which have the same initial conditions but which also demonstrate additional feedback effects. M6 models the effect of SG SNeII development but no AGB winds, leading to a more extended population of SG stars, while M7 shows the effects of SG stars forming SNeII and models AGB feedback as well. The FG of stars in both models appear very similar, but the SG of stars in M7 is less massive and more compact, suggesting greater gas loss.

M10 does not develop a second stellar generation. This is due to the significantly lower mass of the initial SGMC compared to other models ($M_{\rm SGMC} = 10^{6} M_{\odot}$). The resulting star cluster has $R_{\rm eff}\approx18$ pc and is composed solely of FG stars. Star formation in M10 is limited to a single burst in the first 0.5 Myrs, and some of this initial gas persists until the final timestep ($T=298$ Myr). AGB gas is also present by this time, distributed in a loose halo around the stellar cluster, but at no point during the simulation does it collect in densities sufficient to commence further star formation. The low mass of the cluster ($\sim9\times10^{5} M_{\odot}$) is insufficient to prevent gas from being scattered by AGB wind and therefore unable to collect in sufficient density to commence star formation within the time frame of the simulation. However, the AGB gas present in the final timestep suggests that given sufficient time a SG of stars will form.

M14 also shows unusual morphology. The SGMC is low mass ($M_{\rm SGMC}=5\times10^{7} M_{\odot}$) but compact ($R_{\rm SGMC}=45$ pc). Initial star formation proceeds as normal, but SG star formation occurs only slowly for reasons similar to M11: feedback effects strongly counter AGB gas infall, limiting the star formation rate, and by the final timestep the remaining AGB gas is widely distributed. The SG of stars is very low mass ($\approx 2.4\times10^{5} M_{\odot}$) and compact ($R_{\rm eff}\approx0.1 pc$). The FG of stars has become gravitationally unbound as a result of strong feedback effects.

\begin{figure*}
	\includegraphics[width=\linewidth]{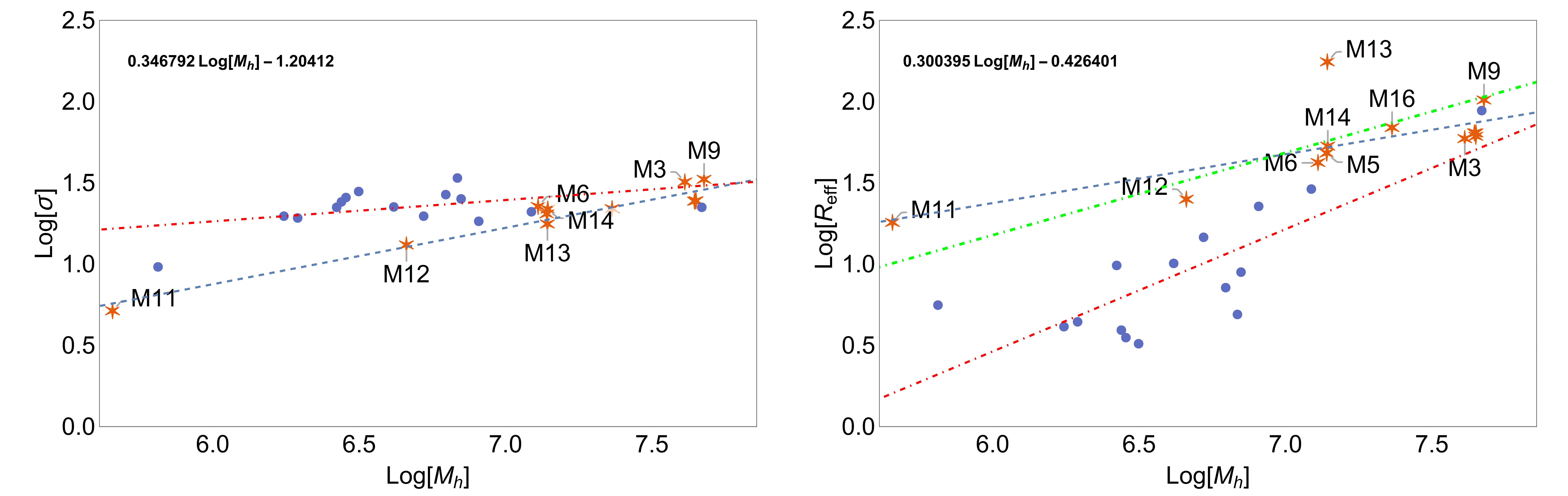}
	\caption{Comparison of all models in final timestep ($T=298$ Myr) with observational data from M08. The simulations are shown as orange stars; the data from M08 are blue points. The dashed blue line is a least-squares regression of all models, the dot-dashed red line is a least-squares regression of the observational points from M08. (\textit{Left}): Average $\sigma_{v}$ (in km/s) plotted against the half-mass (in $M_{\odot}$). (\textit{Right}): $R_{eff}$ (in pc) plotted against half-mass (in $M_{\odot}$). The effective radius for the simulated galaxies is taken as the two-dimensional radius containing half the total mass of all stars. M14 has been excluded from these plots due to an unusually distributed older stellar population. A fit of the relationship between M and R in the original SGMCs is also shown as green dot-dashed line in the right hand pane.}
	\label{fig:comps}
\end{figure*}

Comparisons of the sizes and velocity distributions of the models with observational data from \cite{mieske2008nature} (hereafter M08) are shown in Fig. \ref{fig:comps}. There is a loose linear dependence between $M$ and $R_{\rm eff}$ similar to that which has been observed in UCDs. We use a least squares linear regression to find a relationship of the form
\begin{equation}
 \rm log(R_{\rm eff}) = \textrm{a log}(\rm M_{\odot}) + \rm b
\end{equation}
with parameters $a = 0.30$ and $b = -0.43$. The simulated SCs are more massive and have greater radii than observed UCDs; these are objects that might then undergo tidal stripping \citep{bekki2001galaxy}, resulting in mass loss and a decrease in $R_{\rm eff}$. We expect that the higher mass models would migrate below the line shown in Fig.\ref{fig:comps}, falling more in line with the observational data. The M-R relations in both models and observations appear to converge at high masses. This is most likely a result of lower mass nuclear clusters being less able to retain mass during tidal stripping. The presence of observed UCDs of masses similar to the highest mass models suggests a further population of higher mass star clusters beyond the present models. The simulated SCs also show a close scaling relationships to those of the original SGMCs. 

The relationship between $\sigma_{v}$ and $M$ is also found in the form
\begin{equation}
\rm log(\sigma_{v}) = \textrm{a log}(\rm M_{\odot}) + \rm b
\end{equation}
also shown in Fig. \ref{fig:comps}, with parameters $a = 0.35$ and $b = - 1.20$. The slope of the relation between $\sigma_{v}$ and $M_{h}$ is steeper in our models than in the observational data, however, the scatter appears to be consistent with that observed in UCDs and GCs \citep{Norris2014}.

\section{Discussion}

\subsection{Multiple stellar populations in UCDs}

If UCDs can be formed by the threshing of nucleated elliptical dwarf galaxies (dE,N) then we seek to investigate nucleus formation from SGMC within dwarf galaxies and thereby gain insight into the likely photometric properties of UCDs at their formation. We have conducted sixteen simulations with differing initial parameters and feedback from SNe and AGB stars to investigate the early history of pre-UCD nucleated galaxies. The massive stellar clusters modeled in this investigation have similar physical properties to those observed in UCDs, and all but two of the models show multiple stellar populations. The younger SG of stars more centrally concentrated, at least during the formation phase. We speculate then that UCDs should similarly contain multiple nested populations. Observations of very bright UCDs in the Fornax and Virgo clusters by \cite{evstigneeva2008structural} showed small positive colour gradients, with a distinct population of older red stars surrounding the inner, bluer stars. Our models show no development of metallicity over time, due to the SN ejecta not being retained in the cluster. However, as the SG stars form from AGB ejecta, we expect this population to be rich in He, which may then appear bluer than the original, SGMC-formed stars. This combined with the highly clustered nature of the SG stars could produce something like the observed small colour gradients.

SG star formation history in all models is extensive, which may be a key distinguishing feature for determining the formation mechanism of a UCD. The simulations also show that UCDs formed by this path are similar to GCs. While GCs are typically understood to be dominated by a single old stellar population ($10-12$ Gyr), evidence for the presence of multiple populations within GCs has been observed in GCs in our Galaxy \citep{freeman1975chemical, cohen1981abundances, gratton2001and,bedin2004omega,norris2004helium,piotto2005metallicities,mackey2007double,lee2009chemical,da2014ngc}. 

The star formation histories of resolvable UCDs such as NGC 4546-UCD1 can be very extended \citep{norris2015extended}. Alpha element abundances [$\alpha$/Fe] are an indication of the length of star formation history, with sub-solar abundances pointing to longer star formation timescales and observed UCDs have shown a broad range in these abundances, indicating that at least some have extensive star formation histories \citep{da2011two}, although these abundances are also a function of environment \citep{mieske2007search}.

No model exhibits line of sight velocity of more than a few ($<$10 km s$^{-1}$) km s$^{-1}$. This is typical of GCs and also several observed UCDs such as UCD3 in the Fornax cluster \citep{frank2011spatially}. There are UCDs that have been observed to rotate rapidly, as in the case of M60-UCD1 \citep{seth2014supermassive}, which rotates with $v_{r}\sim40 $ km s$^{-1}$. However, this case is remarkably dense and is host to a supermassive black hole, while those with lower rotation amplitudes may not. The velocity dispersions of the models, ranging from $\sigma_{v}\approx10-30$ km s$^{-1}$ agree with the observations of UCDs in Fornax \citep{mieske2008nature, mieske2012specific, drinkwater2003class}, Virgo \cite{hilker2009ucds}, and Perseus \citep{penny2014ultracompact}, as well as individual UCDs such as NGC 4546-UCD1 \citep{norris2015extended}.

\subsection{Formation of SGMC}

It has been understood for some time that SGMCs are required for the formation of massive star clusters and GCs at high redshifts ($z > 2$) \citep{harris1994supergiant, gnedin2004first, pudritz2002cosmological, Bekki2017a}. In particular, these simulations have suggested that the most massive GCs would form within SGMCs with masses of around $10^{9} M_{\odot}$ which would have been present at very high redshifts, with massive GC formation as early as $z=12$ in some simulations \citep{gnedin2004first} and likely around $z\approx5$ \cite{harris1994supergiant}. As UCDs share many properties with massive GCs, we have here assumed that the massive nucleated star clusters that are UCD progenitors would form under similar circumstances from isolated SGMCs \citep{harris1994supergiant}. 

However, GMCs are unlikely to be long-lived structures: they are typically associated with regions of intense star formation activity, and the associated processes would likely limit the lifespan of a GMC to $\lesssim 3\times10^{7}$ years \citep{blitz1980origin, krumholz2006global}. The delay between formation of a GMC and star formation commencing would not be long, so the lifespans of GMCs will be at most twice the star formation history within them. Similarly, GMCs tend to be closely linked to open clusters, and their masses scale negatively with the masses of the open cluster with which they are associated, so the age of these clusters is an upper limit on the age of those GMCs \citep{leisawitz1989co, larson1981turbulence}. 

Giant molecular clouds could form through the collision and combination of smaller clouds - the "bottom-up" model. This process would take at least $\sim200$ Myr to reach masses that have been observed, in which time star formation activity would have already dispersed them \citep{blitz1980origin}. Collisions between molecular clouds may still contribute to the mass growth of GMCs, but it is not likely to be the chief mechanism. An alternative model proposes the condensation of gas to a region due to magnetohydrodynamic instability such as the Parker instability \citep{blitz1980origin} or a magneto-Jeans instability \citep{kim2002formation}. It is likely that the growth of GMCs is governed by multiple processes dependent on the formation environment.

For super giant molecular clouds, such as the ones modeled in this paper, being very old and considerably more massive, these formation mechanisms are not completely sufficient. SGMCs of the masses required for our model have been observed in the Antennae system \citep{wilson2000high}. Given that such objects appear to be required for the formation of GCs, it has been suggested that self-gravitating massive molecular clouds could form through agglomeration and coalescence in dark matter haloes within a cold dark matter cosmological model \citep{harris1994supergiant, mclaughlin1996model}. \cite{weil2001cosmological} simulated the growth of molecular clouds at very high redshifts to $z=0$ and discovered that this "bottom-up" model of SGMC growth produced clouds of the required mass ($M=10^{7}-10^{8} M_{\odot}$) and greater, and suggests that such objects would reach SGMC status by $z\approx8$. SGMCs could conceivably form in this way during the very gas-rich phases of dwarf galaxies at high z, or by accumulation of gas through dwarf-dwarf interaction and merging. The gas channeling action of bar formation could also give rise to large MCs which could then combine to produce SGMCs.

Should one or several SGMC form in a non-nucleated dwarf galaxy and generate massive star clusters of the type simulated in this investigation, we expect that the SGMC would not form centrally within the dwarf galaxy. The timescale of dynamical friction in dwarf galaxies has been discussed, using members of the local group, Fornax and Virgo as references \citep{tremaine1976formation, oh2000globular, hernandez1998dynamical, lotz2001dynamical}. The time taken for a cluster of mass M at initial radius $r_{i}$ within an isothermal halo, moving with circular velocity $v_{c}$, to reach the centre is given in \cite{binney1987galactic} (BT87):

\begin{equation}
t_{DF} = \frac{0.529}{\textrm{ln } \Lambda}(\frac{r_{i}}{500 \textrm{ pc}})^{2}(\frac{v_{c}}{80 \textrm{ km s}^{-1}})(\frac{10^{7} M_{\odot}}{\rm M}) \rm{ Gyr}
\end{equation}

The full derivation is available in (BT87). Here ln $\Lambda$ is the Coulomb logarithm, defined by the impact parameter b$_{\rm{max}}$, the system velocity V$\sim v_{c}$ and star mass m:

\begin{equation}
\textrm{ln }\Lambda = \rm{ln}(\frac{\rm{b}_{\rm{max}}\rm{V}^{2}}{\rm G \rm M})
\end{equation}

We assume here that an elliptical dwarf galaxy is adequately modeled by an isothermic profile and assume ln $\Lambda$ = 3 \citep{bekki2005formation}. The mass of the stellar cluster in the fiducial model is M$_{sc} \approx 10^{7} M_{\odot}$, and we assume an initial radius of the cluster $r_{i} = 500 \rm{ pc}$. We expect then that $t_{DF}\approx0.2$ Gyr. Cluster masses greater than this will decay into the central region in far shorter timescales, $t_{DF}<0.1$ Gyr. This model of dynamical friction, which describes dynamical friction in a dark matter halo, will likely overestimate the time taken for the SC to move into the center of the galaxy. For a more accurate estimation of $t_{DF}$ we need to formulate a model for dynamical friction of a SC in a disk galaxy. However, based on previous work in \cite{bekki2010dynamical} suggesting the dynamical friction timescales of typical GCs with masses $M_{gc} = 2\times10^{5} \rm M_{\odot}$ is on the order of 1 Gyr, our estimation that the more massive SCs would have $t_{DF}$ of 0.1 Gyr or less is reasonable. 

There are other mechanisms by which a dwarf galaxy could become nucleated without such massive star clusters. The interaction between a supermassive black hole (SMBH) at the centre of the galaxy with GCs decreases the dynamical friction timescale dramatically and could lead to the development of a nucleus within 2 Gyr \citep{arca2017megan}. Dissipative merging of smaller star clusters and gaseous clumps in gaseous spiral arms can transfer a significant percentage of the host mass ($\sim 5\%$) to the core to form a nucleus without requiring the existence of an SGMC \citep{bekki2007formation}. Nuclear star clusters of sizes comparable to those modeled in this investigation have been observed and may be very common; \cite{georgiev2014nuclear} catalogued 228 such clusters in nearby spiral galaxies.

The present investigation does not include the effects of the wider environment, i.e. the dwarf galaxy in which the massive SC is formed. The study by \cite{Guillard2016} investigates a similar scenario of massive nuclear SC formation within the broader context of a dwarf galaxy, however, there are several distinctions between that work and this investigation. The spatial and mass resolution of our models are far higher than that of the highest resolution simulation run by \cite{Guillard2016}, and the present work is more focused on the formation and internal dynamics of a single, very massive nuclear star cluster from a single SGMC, while \cite{Guillard2016} shows the development and merger of several less massive clusters. The difference in the dynamical friction timescale is accounted for by the lower initial masses of the clusters in \cite{Guillard2016}. The tidal effect on the sort of massive SC formation from SGMCs investigated here is included in B17, however, it is only to strip the outermost gas from the cluster and is not considered a large effect on clusters or SGMCs at this scale. Further investigation into our scenario including the broader context of the surrounding dwarf galaxy would certainly provide greater insight, however, for our purposes here the internal dynamics were of greater importance.

\subsection{Future work}

We have shown that a single SGMC within a non-nucleated dwarf galaxy might become a massive star cluster with the potential to first become the nucleus of the dwarf galaxy and subsequently be threshed, thereby becoming a UCD. From these 15 models, it seems likely that there could be a population of these proto-UCD stellar clusters more massive than these; UCDs of similar mass and size to our models have been observed and we expect that these clusters would lose some size and mass during threshing. So an investigation into the evolution of SGMCs with initial mass $ > 10^{8} M_{\odot}$ would be beneficial; whether such objects would form, or if this population of massive SCs must be formed through mergers of SCs of a size similar to those modeled here. Lower-mass models showed signs of ongoing SG star formation, and continued models of these clusters in the timescale range T = 300 Myr - 1 Gyr may show further development of the SCs beyond these simulations. In particular, where feedback from SNe and AGB winds may be likely to eject AGB gas from the cluster, simulations at these timescales may help establish a lower bound for proto-UCD SC masses. Further investigation into the formation of SGMC within non-nucleated dwarf galaxies would be highly beneficial for establishing the validity of this evolution pathway. A more thorough investigation into $t_{DF}$ of SCs of this mass in disk galaxies by dynamical simulation would also help establish how realistic the present scenario is,\ textbf{as would further investigation including the broader context of a dwarf galaxy}. 

\section{Conclusions}

We have simulated the formation of very massive star clusters from SGMCs in dwarf galaxies in order to understand the origin of UCDs. In the scenario explored here, these clusters can be later evolved into massive nuclei of dwarf galaxies, which may then become UCDs through threshing by proximity to another massive galaxy. The principal results of these simulations are summarised here:

\begin{enumerate}
	\item Collapsing SGMCs demonstrate initial violent star formation, giving way to slower, ongoing star formation beginning at around $T\approx30$ Myr and continuing until the end of the simulation through the collection of AGB-emitted gas particles. For SGMC of mass $M_{\rm mc} \geq 10^7 M_{\odot}$, this results in more than one distinct stellar population.
	\item The stellar populations formed are highly nested; the FG stars form an extended halo; in the fiducial model, the FG of stars has an effective radius $R_{\rm eff}\approx58$ pc while later generations are highly constrained with $R_{eff}\approx18$ pc. The centre of mass of later generations is not necessarily aligned with that of the FG. \item The younger stellar populations will be He-rich and therefore may appear bluer than those stars formed from SGMC gas. This, combined with the highly clustered nature of the younger populations, may produce a small colour gradient from a bluer, He-rich core to a less blue FG envelope.
	\item Feedback from SNe and AGB wind primarily influences the rate at which star formation proceeds. However, for more compact and less massive SGMCs, the feedback effects may overcome the gravitational constraints on the FG of stars, leading to non-elliptical distributions while SG star formation proceeds much more slowly than in more massive models. 
	\item The physical properties of the SCs are similar to those observed in UCDs, including: (a) evidence of multiple stellar populations; (b) extended star formation histories; (c) masses; (d) effective radii; (e) $\sigma_{v}$ and rotation, and (f) scaling relations. In particular, the SCs exhibit M-R scaling relations consistent with observed UCDs. These scaling relations also strongly follow those of the SGMCs from which the clusters formed.
	\item Preliminary estimates of dynamical friction timescale indicate that the stellar cluster could migrate to the centre of the dwarf galaxy in $<$ 1 Gyr, so our scenario is quite realistic. 	
\end{enumerate}

We have discussed how massive stellar clusters formed from SGMCs could become the nuclei of dwarf galaxies and subsequently, through a tidal threshing process, become UCDs. These results, while preliminary, are a good starting point for further investigation into this topic, which may provide an understanding of the photometric properties of UCDs early in their lifetimes. We have also discussed how SGMC could form in dwarf galaxies, and suggest that ongoing exploration of how these might form in dwarf galaxies would be beneficial for future work on the topic of UCD formation.

\section{Acknowledgements}

We would like to thank the referee of this paper for many constructive and useful comments that have greatly improved it.

\bibliographystyle{mnras}
\bibliography{ucdn}

\label{lastpage}

\end{document}